\documentclass[a4paper,11pt]{article}
\usepackage[DIV12]{typearea}
\usepackage{longtable}
\usepackage{booktabs,colortbl}
\usepackage{multirow}
\usepackage{amsmath}
\usepackage{amssymb}
\usepackage{bbm}
\usepackage[utf8]{inputenc}
\usepackage{pdflscape}
\usepackage{xspace}
\usepackage[caption=false]{subfig}
\usepackage{nicefrac}
\usepackage{graphicx}
\usepackage{cite}  
\usepackage[small]{caption2}
\usepackage{scalerel} %to scale subscripts in equations
\usepackage{mathtools}

\usepackage{adjustbox}%\adjincludegraphics

\numberwithin{equation}{section}

\newcommand{\ttitle}{Supersymmetric brick wall diagrams and the dynamical fishnet}

\newcommand{\SU}[1]{\ensuremath{\mathrm{SU}(#1)}}

\newcommand{\e}{\mathrm{e}}
\newcommand{\I}{\mathrm{i}}

\newcommand{\com}[2]{\lbrack #1, #2\rbrack}

\newcommand{\x}{\ensuremath{\times}}

\newcommand{\dd}{\mathrm{d}}

\usepackage{xcolor}

\advance \headheight by 3.0truept       % for 12pt mandatory...
\usepackage{hyperref}
\hypersetup{
    pdftitle = {\ttitle},
    pdfauthor = {Kade, Staudacher}
}

\begin{document}

\begin{titlepage}

%\vspace*{-3.0cm}
\begin{flushright}
\normalsize{HU-EP-24/22-RTG}\\
\end{flushright}

\vspace*{1.0cm}

\begin{center}
{\Large\textbf{\boldmath \ttitle}\unboldmath}

\vspace{1cm}

\textbf{Moritz Kade}, \textbf{Matthias Staudacher}
\\[4mm]
\begin{small}
\texttt{\string{\href{mailto:mkade@physik.hu-berlin.de}{mkade},\href{mailto:staudacher@physik.hu-berlin.de}{staudacher}\string}@physik.hu-berlin.de}
\end{small}
\\[8mm]
\textit{\small Institut f\"{u}r Mathematik und Institut f\"{u}r Physik,\\
Humboldt-Universit\"{a}t zu Berlin,\\
Zum Großen Windkanal 2, 12489 Berlin, Germany}
\end{center}

\vspace{1cm}

\vspace*{1.0cm}

\begin{abstract}
We consider the double scaling limit of $\beta$-deformed planar $\mathcal{N}=4$ supersymmetric Yang-Mills theory (SYM), which has been argued to be conformal and integrable. It is a special point in the three-parameter space of double-scaled $\gamma_i$-deformed $\mathcal{N}=4$ SYM,  preserving $\mathcal{N}=1$ supersymmetry. The Feynman diagrams of the general three-parameter models form a ``dynamical fishnet'' that is much harder to analyze than the original one-parameter fishnet, where major progress in uncovering the model's integrable structure has been made in recent years. Here we show that by applying $\mathcal{N}=1$ superspace techniques to the \mbox{$\beta$-deformed} model the dynamical nature of its Feynman graph expansion disappears, and we recover a regular lattice structure of brick wall (honeycomb) type. As a first application, we compute the zero-mode-fixed thermodynamic free energy of this model by applying Zamolodchikov's method of inversion to the supersymmetric brick wall diagrams.
\end{abstract}

\end{titlepage}

\newpage

%%%%%%%%%%%%%%%%%%%%%%%%%%%%%%%%%%%%%%%%%%%%%%%%%%%%%%%%%%%%%%%%%%%%%%%%%%%%%%%%%%%%%%%%%%%%%%%%%%%%%%%%%%%%%%%%%%%%%%%%%%%%%%%%%%%%%%%%%
\section{Introduction and results}
\label{sec:Introduction}
As a quantum field theory (QFT), $\mathcal{N}=4$ supersymmetric Yang-Mills theory in the planar limit (SYM) is historically called integrable because one is able to map the problem of finding anomalous dimensions of operators, up to a certain loop order, to the spectral problem of a suitable integrable spin chain Hamiltonian \cite{Beisert:2010jr}.
Apart from techniques based on other incarnations of integrability \cite{Bombardelli:2016rwb,Gromov:2011cx,Gromov:2014caa,Gromov:2017blm,Basso:2013vsa,Basso:2014hfa}, which can access much higher orders in perturbation theory, the task of determining the Hamiltonian becomes increasingly hard at higher loop orders due to the many Feynman diagrams contributing to two-point functions of local composite operators.
The theory's large number of fields and the many ways for them to interact are to blame for the rapidly increasing complexity.

R-symmetry deformed versions of $\mathcal{N}=4$ SYM in the so-called double-scaling limit \cite{Gurdogan:2015csr} are much simpler as regards their diagrammatics, mostly because the gauge degrees of freedom decouple.
The double-scaling limit consists in increasing the three complexified deformation parameters while decreasing the 't Hooft coupling, balanced in such a way that their product stays finite.
As a further consequence, only interaction terms with this particular combination of the three parameters survive this limit \cite{Gurdogan:2015csr}. 
Further simplifications and decouplings are possible by reducing the parameters from three to two or even only one \cite{Caetano:2016ydc}.
All these models are non-unitary, because the hermitian conjugated counter parts of the interaction terms do not survive the limit.

The one-parameter bi-scalar fishnet theory \cite{Gurdogan:2015csr}, diagrammatically discovered as early as 1980 \cite{Zamolodchikov:1980mb}, is an example where only one interaction term survives.
The theory's Feynman diagrams possess a very regular lattice (fishnet) structure and contain only scalar propagators.
Many investigations have shown that here its integrability is based on the so-called star-triangle relation (STR) \cite{DEramo:1971hnd,Symanzik:1972wj,UniquenessDIKazakov}, a Feynman diagram relation between a star- and a triangle-shaped graph, which implies the existence of an R-matrix satisfying a Yang-Baxter equation \cite{Chicherin:2012yn}.
This insight allows to construct commuting transfer matrices, which are subgraphs of the theory's Feynman diagrams, generalized by a (spectral) parameter in the propagators' exponents.
The existence of a STR gives a new perspective on integrability: a duality with integrable, statistical lattice models \cite{Au-Yang:1999kid,Bazhanov:2016ajm}.
Therein, a Feynman diagram corresponds to a partition function of the associated lattice model and the Boltzmann weights correspond to the scalar propagators. 
The observable to which the diagram contributes to in the QFT dictates the boundary conditions of the related partition function.

From the point of view of the lattice model, one natural choice are periodic boundary conditions.
These nicely agree with the leading large-$\mathrm{N}$ vacuum diagrams of the fishnet CFT, which are indeed of toroidal topology due to their regular, flat lattice structure.
A quantity of high interest in the field of integrable lattice models is the free energy in the thermodynamic limit, whose QFT analog is the critical coupling of the theory, i.\ e.\ the radius of convergence of the free energy's perturbative expansion.
This computation was first done by Zamolodchikov \cite{Zamolodchikov:1980mb} using the method of inversion relations \cite{StroganovInvRelLatticeModels,Baxter:1982xp,Baxter:1989ct,Baxter:2002qg,Pokrovsky_1982,bousquet1999inversion}.
As such, it was the first analytic result for the fishnet model, even before the Lagrangian producing the fishnet graphs was written down \cite{Gurdogan:2015csr}.

Despite much progress in understanding and applying the integrable structure of fishnet theory, the eventual goal must be to return back to its much more intricate ``mother theory'', full-fledged $\mathcal{N}=4$ SYM.
An important stopover along this challenging route are the above mentioned three-parameter double-scaled theories, the so-called $\chi$-CFTs \cite{Gurdogan:2015csr,Caetano:2016ydc,Kazakov:2018gcy}.
Their field content is already much closer to $\mathcal{N}=4$ SYM, merely lacking the gauge fields and one of the four fermions.
When trying to make the connection to a lattice model, one notices the apparent non-existence of a suitable transfer matrix, due to the simultaneous presence of cubic Yukawa-type vertices and quartic bosonic vertices.
It has been suggested in \cite{Kazakov:2018gcy} that one rather needs a ``dynamical'' object, where the admixture of cubic and quartic vertices is somehow automatically generated.
However, no concrete construction has been proposed, up to now.

In the present paper, we propose just such a construction in the special case of the $\mathcal{N}=1$ supersymmetric $\chi$-CFT.
To be precise, we employ $\mathcal{N} = 1$ Feynman supergraphs in order to study the double-scaling limit of the $\mathcal{N}=1$ supersymmetric $\beta$-deformation of $\mathcal{N}=4$ SYM.
In close similarity to the $\mathcal{N} = 0$ fishnet theory, this nicely homogenizes the occurring perturbative diagrams to a regular brick wall structure.
First, we show that the double-scaled $\mathcal{N} = 1$ superspace action reproduces the known supersymmetric $\chi$-CFT from the literature \cite{Gurdogan:2015csr,Caetano:2016ydc}.
We then derive generalized superfield propagators containing a spectral parameter, leading to our proposal for the weights of a corresponding lattice model.
In analogy with \cite{Kazakov:2018qbr,Alfimov:2023vev}, we propose a suitable non-local action that includes a spectral parameter.
Excitingly, it still appears to be formally supersymmetric.
We then derive superconformal integral relations from a superspace star integral due to Osborn \cite{Osborn:1998qu}.
Mysteriously, it slightly falls short of being a proper star-triangle relation, while still allowing us to suitably adapt Zamolodchikov's calculation of the model's critical coupling by inversion relations. 
His extremely concise computation was reproduced in detail in our earlier work \cite{Kade:2023xet}, where we applied it to the free energy of a fermionic brick wall model.
In fact, the present work may be considered to be a supersymmetric generalization of this article.
Thereby, we are able to find the exact critical coupling of the $\mathcal{N}=1$ double-scaled $\beta$-deformation of $\mathcal{N}=4$ SYM along with its spectral deformation.

%%%%%%%%%%%%%%%%%%%%%%%%%%%%%%%%%%%%%%%%%%%%%%%%%%%%%%%%%%%%%%%%%%%%%%%%%%%%%%%%%%%%%%%%%%%%%%%%%%%%%%%%%%%%%%%%%%%%%%%%%%%%%%%%%%%%%%%%%
\section{\boldmath The double-scaled \texorpdfstring{$\beta$}{}-deformation of \texorpdfstring{$\mathcal{N}=4$}{} SYM}
$\mathcal{N} = 4$ super Yang-Mills theory (SYM) in the large $\mathrm{N} \rightarrow \infty$ limit can be conveniently formulated in the $\mathcal{N} = 1$ superspace description with the action \cite{Penati:2001sv}
\begin{equation}
\begin{split}
S
= &
\int \dd^4 x\; \dd^2\theta \dd^2 \bar{\theta}\;
\sum_{i=1}^{3} \,
\mathrm{tr} \left[ 
\e^{-g V} \Phi_i^\dagger \e^{g V} \Phi_i
\right]
+
\frac{1}{2g^2}
\int \dd^4 x\; \dd^2\theta\;
\mathrm{tr} \left[
W^\alpha W_\alpha
\right]\\
& +
\I g
\int \dd^4 x\; \dd^2\theta\;
\mathrm{tr} \left[
\Phi_1 \com{\Phi_2}{\Phi_3}
\right]
+
\I g
\int \dd^4 x\; \dd^2\bar{\theta}\;
\mathrm{tr} \left[
\Phi^\dagger_1 \com{\Phi^\dagger_2}{\Phi^\dagger_3}
\right] ~.
\end{split}
\label{eq:SYM_Action}
\end{equation}
It comprises a real vector superfield $V$ and three chiral superfields $\Phi_i$, all four in the adjoint representation of the gauge group $\SU{\mathrm{N}}$, i.\ e.\ $V = V_A T^A$ and $\Phi_i = \Phi_{i,A} T^A$. The shorthand $W_\alpha = \I \bar{D}^2 \left( \e^{-g V} D_\alpha \e^{g V} \right)$ is used, with the covariant superderivatives defined in \eqref{eq:SuperCovariantDerivatives}.

Next, we would like to deform the $\SU{4}$ R-symmetry of \eqref{eq:SYM_Action} by replacing the ordinary products of fields with the star product $\Phi_i \star \Phi_j := \e^{\I\; \mathrm{det}\left(\boldsymbol{\gamma} \vert \mathbf{q}_i \vert \mathbf{q}_j\right)} \Phi_i \Phi_j$ and setting the three deformation parameters to the same value $\boldsymbol{\gamma} = (\beta, \beta, \beta)$ \cite{Imeroni:2008cr}.
The $\mathbf{q}_i$ are the charges of $\Phi_i$ under the Cartan elements of the R-symmetry $\SU{4}$.
They are $\mathbf{q}_1 = \left( \frac{1}{2}, -\frac{1}{2}, -\frac{1}{2} \right)$, $\mathbf{q}_2 = \left( -\frac{1}{2}, \frac{1}{2}, -\frac{1}{2} \right)$ and $\mathbf{q}_3 = \left( -\frac{1}{2}, -\frac{1}{2}, \frac{1}{2} \right)$ for the three chiral superfields and zero for the gauge field \cite{Jin:2012np}.
This deformation leaves us with the superpotential of the $\beta$-deformation of $\mathcal{N}=4$ SYM \cite{Lunin:2005jy,Jin:2012np,Fokken:2013mza}
\begin{equation}
\I g \int \dd^4 x \; \dd^2 \theta ~ \mathrm{tr}\left[ q\; \Phi_1 \Phi_2 \Phi_3 - q^{-1} \Phi_1 \Phi_3 \Phi_2 \right] + \mathrm{h.c.} ~,
\label{eq:MarginalDeformation}
\end{equation}
where we introduced the abbreviation $q = \e^{\I\beta}$.

Next, one performs the 't Hooft limit, by sending $g\rightarrow 0$ and $\mathrm{N} \rightarrow \infty$, while keeping the 't Hooft coupling $\lambda := g^2 \mathrm{N}$ fixed.
After appropriate rescalings of the action and the fields, Feynman graphs in double-line notation with the smallest genus will dominate.
In a second step, we perform the double-scaling limit consisting of $\lambda \rightarrow 0$ and $\beta \rightarrow -\I \infty \Rightarrow q\rightarrow \infty$, while the product $\xi := \lambda\cdot q$ remains finite. 
After this operation the gauge fields decouple and two out of the four terms of the deformation \eqref{eq:MarginalDeformation} vanish. 
We consider the obtained theory in the planar limit $\mathrm{N}\rightarrow \infty$, where the leading order is described by toroidal double-line Feynman graphs.
We obtain the concise action
\begin{equation}
\begin{split}
S ~=~
& S_\mathrm{kin} + S_\mathrm{int}\\
~=~
& \mathrm{N}\int \dd^4 x\; \dd^2\theta \dd^2 \bar{\theta}
\left\lbrace 
\sum_{i=1}^{3} 
\mathrm{tr} \left[ 
\Phi_i^\dagger \Phi_i
\right]
+
\I \xi \cdot \bar{\theta}^2\,
\mathrm{tr}\left[ \Phi_1 \Phi_2 \Phi_3 \right]
+
\I \xi \cdot \theta^2\,
\mathrm{tr}\left[ \Phi_1^\dagger \Phi_2^\dagger \Phi_3^\dagger \right]
\right\rbrace ~,
\end{split}
\label{eq:DSbetaDeformation}
\end{equation}
where the squares $\theta^2$ and $\bar{\theta}^2$ in general act as delta function in the fermionic coordinates, see appendix \ref{sec:BerezinIntegral}.
Accordingly, we can read off the two Feynman supergraph vertices in Minkowski space
\begin{equation}
\adjincludegraphics[valign=c,scale=0.5]{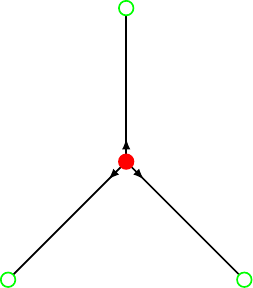}
\sim
- \xi \int \dd^4x\; \dd^2\theta\,\dd^2\bar{\theta} \;\delta^{(2)}(\bar{\theta}) ~, ~~~
\adjincludegraphics[valign=c,scale=0.5]{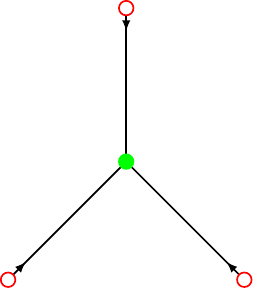}
\sim
- \xi \int \dd^4x\; \dd^2\theta\,\dd^2\bar{\theta} \;\delta^{(2)}(\theta) ~.
\label{eq:betaDef_FeynmanRules_vertices}
\end{equation}
Here and in the following we will denote internal, integrated, chiral (anti-chiral) vertices by a filled red (green) dot, corresponding to the left (right) vertex in \eqref{eq:betaDef_FeynmanRules_vertices}.
Note that the Gra\ss mann delta functions in \eqref{eq:betaDef_FeynmanRules_vertices} annihilate the part of the fermionic integration whose chirality is opposite to the one of the vertex at hand.
Hence, propagators always connect the chiral and anti-chiral subspaces of superspace.
When an external point in a super Feynman diagram is expected to be integrated by a chiral (anti-chiral) vertex in a later step according to the Feynman vertex rules \eqref{eq:betaDef_FeynmanRules_vertices}, we denote this by a un-filled red (green) circle.

\subsection{Component action}
Let us now make contact with the component field action of the $\mathcal{N} = 1 $ double-scaled $\beta$-deformation of $\mathcal{N}=4$ SYM.
The three chiral and anti-chiral superfields possess the following component expansions
\begin{subequations}
\begin{align}
\Phi_i 
&=
\phi_i (x_+) + \sqrt{2}\; \theta \psi_i (x_+) + \theta^2 F_i(x_+)
=
\e^{\I \theta \sigma^\mu \bar{\theta} \partial_\mu}
\left[
\phi_i (x) + \sqrt{2}\; \theta \psi_i (x) + \theta^2 F_i(x)
\right]\\
&=
\phi_i (x)  +  \I \theta \sigma^\mu \bar{\theta} \partial_\mu \phi_i (x)  +   \frac{1}{4} \theta^2 \bar{\theta}^2 \square \phi_i (x)  +  \sqrt{2}\; \theta \psi_i (x)  -  \frac{\I}{2} \theta^2 \partial_\mu \psi_i(x) \sigma^\mu \bar{\theta}  +  \theta^2 F_i (x)\\
\Phi^\dagger_i 
&=
\phi^\dagger_i (x_-) + \sqrt{2}\; \bar{\theta} \bar{\psi}_i (x_-) + \bar{\theta}^2 F^\dagger_i(x_-)
=
\e^{-\I \theta \sigma^\mu \bar{\theta} \partial_\mu}
\left[
\phi^\dagger_i (x) + \sqrt{2}\; \bar{\theta} \bar{\psi}_i (x) + \bar{\theta}^2 F^\dagger_i(x)
\right]\\
&=
\phi^\dagger_i (x)  -  \I \theta \sigma^\mu \bar{\theta} \partial_\mu \phi^\dagger_i (x)  +   \frac{1}{4} \theta^2 \bar{\theta}^2 \square \phi^\dagger_i (x)  -  \sqrt{2}\; \bar{\theta} \bar{\psi}_i (x)  +  \frac{\I}{2} \bar{\theta}^2 \theta \sigma^\mu \partial_\mu \bar{\psi}_i(x)  +  \bar{\theta}^2 F^\dagger_i (x)
\end{align}\label{eq:Superfield_Components}%
\end{subequations}
with $\square = \partial^\mu \partial_\mu$ and $x^\mu_\pm = x^\mu \pm \I \theta \sigma^\mu \bar{\theta}$. 
Here, the $\phi_i$ denote three ordinary complex scalar fields, the $\psi_i$ the three Weyl fermions and $F_i$ are three non-propagating auxiliary fields.
In appendix \ref{sec:Notations}, the suppression of spinor indices is explained in detail.

We may now expand \eqref{eq:DSbetaDeformation} in Gra\ss mann components.
The kinetic terms from the canonical K\"{a}hler potential are 
\begin{equation}
S_\mathrm{kin} =
\mathrm{N}\int \dd^4 x\; \dd^2\theta \dd^2 \bar{\theta} 
\sum_{i=1}^{3} 
\mathrm{tr} \left[ 
\Phi_i^\dagger \Phi_i
\right]
=
\mathrm{N}\int \dd^4 x\; 
\sum_{i=1}^{3} 
\mathrm{tr} \left[ 
\phi_i^\dagger \square \phi_i -
\I \bar{\psi}_i\bar{\sigma}^\mu \partial_\mu \psi_i +
F_i^\dagger F_i
\right] ~,
\end{equation}
while the interaction terms are
\begin{equation}
\begin{split}
S_\mathrm{int} 
&=
\mathrm{N} \cdot \I \xi 
\int \dd^4 x\; \dd^2\theta \;
\mathrm{tr} \left[
\Phi_1 \Phi_2 \Phi_3
\right]_{\bar{\theta} = 0}
+
\mathrm{N} \cdot \I \xi
\int \dd^4 x\; \dd^2\bar{\theta} \;
\mathrm{tr} \left[
\Phi^\dagger_1 \Phi^\dagger_2 \Phi^\dagger_3
\right]_{\theta = 0}\\
&=
\mathrm{N} \cdot \I \xi 
\int \dd^4 x \;
\mathrm{tr} \left[
\phi_1 \phi_2 F_3 + 
\phi_1 F_2 \phi_3 +
F_1 \phi_2 \phi_3 - 
\phi_1 \psi_2 \psi_3 -
\phi_2 \psi_3 \psi_1 -
\phi_3 \psi_1 \psi_2
\right]\\
& \phantom{=} +
\mathrm{N} \cdot \I \xi
\int \dd^4 x \;
\mathrm{tr} \left[
\phi^\dagger_1 \phi^\dagger_2 F^\dagger_3 + 
\phi^\dagger_1 F^\dagger_2 \phi^\dagger_3 +
F^\dagger_1 \phi^\dagger_2 \phi^\dagger_3 - 
\phi^\dagger_1 \bar{\psi}_2 \bar{\psi}_3 -
\phi^\dagger_2 \bar{\psi}_3 \bar{\psi}_1 -
\phi^\dagger_3 \bar{\psi}_1 \bar{\psi}_2
\right] ~.
\end{split}
\end{equation}
The auxiliary fields $F_i$ and $F^\dagger_i$ are not dynamical and we may eliminate them with the help of their equations of motion:
\begin{equation}
\begin{aligned}[c]
F_1^A = - \I \xi \, \phi^*_{2,B} \phi^*_{3,C} \cdot \mathrm{tr}\left[ T^A T^B T^C \right] ~,\\
F_2^A = - \I \xi \, \phi^*_{3,B} \phi^*_{1,C} \cdot \mathrm{tr}\left[ T^A T^B T^C \right] ~,\\
F_3^A = - \I \xi \, \phi^*_{1,B} \phi^*_{2,C }\cdot \mathrm{tr}\left[ T^A T^B T^C \right] ~,
\end{aligned}
\qquad
\begin{aligned}[c]
F^{*,A}_1 = - \I \xi \, \phi_{2,B} \phi_{3,C} \cdot \mathrm{tr}\left[ T^A T^B T^C \right] ~,\\
F^{*,A}_2 = - \I \xi \, \phi_{3,B} \phi_{1,C} \cdot \mathrm{tr}\left[ T^A T^B T^C \right] ~,\\
F^{*,A}_3 = - \I \xi \, \phi_{1,B} \phi_{2,C} \cdot \mathrm{tr}\left[ T^A T^B T^C \right] ~.
\end{aligned}
\end{equation}
As in \cite{Fokken:2013aea}, making use of the fact that the generators of $\SU{\mathrm{N}}$ obey $\mathrm{tr}\left[ T^A T^B \right] = \delta^{AB}$ and $( T^A )^a_b \left( T_A \right)^c_d = \delta^a_d \delta^c_b - \frac{1}{\mathrm{N}} \delta^a_b \delta^c_d$, one obtains the action on bosonic space
\begin{equation}
\begin{split}
S
=
\mathrm{N}\int \dd^4 x\;  
\mathrm{tr} 
\left\lbrace
\sum_{i=1}^{3}
\left[ 
\phi_i^\dagger \square \phi_i -
\I \bar{\psi}_i\bar{\sigma}^\mu \partial_\mu \psi_i
\right]
+
\xi^2
\left[
\phi_1 \phi_2 \phi^\dagger_1 \phi^\dagger_2 +
\phi_3 \phi_1 \phi^\dagger_3 \phi^\dagger_1 +
\phi_2 \phi_3 \phi^\dagger_2 \phi^\dagger_3
\right]
\right.\\
\left.
- 
\I \xi
\left[
\phi_1 \psi_2 \psi_3 +
\phi_2 \psi_3 \psi_1 +
\phi_3 \psi_1 \psi_2 
\right]
- 
\I \xi
\left[
\phi^\dagger_1 \bar{\psi}_2 \bar{\psi}_3 +
\phi^\dagger_2 \bar{\psi}_3 \bar{\psi}_1 +
\phi^\dagger_3 \bar{\psi}_1 \bar{\psi}_2
\right]
+
\mathcal{L}_\mathrm{dt}
\right\rbrace ~. \label{eq:DS_components}
\end{split}
\end{equation}
Here $\mathcal{L}_\mathrm{dt}$ describes double-trace interaction terms, which survive in the double-scaling limit and read
\begin{equation}
\mathcal{L}_\mathrm{dt}
=
-\frac{\xi^2}{\mathrm{N}}
\left\lbrace
\mathrm{tr}
	\left[
	\phi_1 \phi_2
	\right]
\mathrm{tr}
	\left[
	\phi_1^\dagger \phi_2^\dagger
	\right]
+
\mathrm{tr}
	\left[
	\phi_1 \phi_3
	\right]
\mathrm{tr}
	\left[
	\phi_1^\dagger \phi_3^\dagger
	\right]
+
\mathrm{tr}
	\left[
	\phi_2 \phi_3
	\right]
\mathrm{tr}
	\left[
	\phi_2^\dagger \phi_3^\dagger
	\right]
\right\rbrace ~.
\label{eq:DoubleTraceTerms}
\end{equation}
It is satisfying to see that they do not need to be added ``by hand'' to the single-trace component action; instead they naturally and directly emerge from the $\mathcal{N} = 1$ superspace formulation.
They are however not needed in our below analysis of the model's critical coupling.
One may compare \eqref{eq:DS_components} to the double-scaled $\beta$-deformation of $\mathcal{N} = 4$ SYM, see eq.\ (3) in \cite{Gurdogan:2015csr} (after changing the convention $\xi \rightarrow -\xi$).

%\MK{\cite{Fokken:2013aea,Kazakov:2018gcy}.
%By tuning the double trace couplings to a fixed point of the beta function, one is able to obtain a superconformal theory \cite{Fokken:2013aea,Gurdogan:2015csr}.
%However, the interaction terms \eqref{eq:DoubleTraceTerms} are all of order $\frac{1}{\mathrm{N}}$ in the large $\mathrm{N}$ limit, and since we are interested in large toroidal diagrams we neglect them in the following.
%}

\subsection{\boldmath Introducing a spectral parameter into the \texorpdfstring{$\mathcal{N} = 1$}{} action}
The double-scaled $\chi$-CFTs are expected to inherit integrability from their $\mathcal{N} = 4$ SYM ``parent theory'' \cite{Gurdogan:2015csr}, which should include the special case \eqref{eq:DSbetaDeformation}.
However, putting quantum integrability to good use always requires the introduction of a suitable spectral parameter.
In general, it is not easy to find it in an integrable, planar, conformal QFT.
This is certainly the case for $\mathcal{N} = 4$ SYM, where it somewhat mysteriously first appears when converting local composite operators into quantum spin chains.
Impressively, in the much simpler setting of the fishnet model, it has recently been shown in \cite{Kazakov:2018qbr} that the correct spectral parameter may be directly introduced by deforming the model's action, at the cost of giving up the locality of the resulting ``QFT''. 
Unfortunately, this construction has not yet been achieved for general $\chi$-CFTs.
We will now show that in the special case of the $\mathcal{N} = 1$ $\chi$-CFT, a suitable deformation may nevertheless be found by deforming its $\mathcal{N} = 1$ action.
%Integrability requires to widen the scope to the theory space containing the object of study, in our case \eqref{eq:DSbetaDeformation}. 
%This integrable fiber is parameterized by the spectral parameter and for a fixed value we find our theory under investigation.
%However, it is also interesting to consider theories at other points of the spectral parameter.
%For us, the spectral parameter will enter in the exponents of the super propagators.
%Once we set the exponent to $1$ we recover the double-scaled $\beta$-deformation \eqref{eq:DSbetaDeformation}.
%In order to make the exponent physical \cite{Alfimov:2023vev}, we have to modify the kinetic terms of the component fields, or in the language of the chiral superfield, we have to modify the canonical K\"{a}hler potential.
%Its diagonal part gets enhanced by an extra d'Alembertian raised to the spectral parameter, for each of the three flavors,
%The deformed action is
We propose
\begin{equation}
\begin{split}
S_{\boldsymbol{\omega}} ~=~
& S_{\mathrm{kin},\boldsymbol{\omega}} + S_{\mathrm{int},\boldsymbol{\omega}}\\
~=~
& \mathrm{N}\int \dd^4 x\; \dd^2\theta \dd^2 \bar{\theta}
\left\lbrace 
\sum_{i=1}^{3} 
\mathrm{tr} \left[ 
\Phi_i^\dagger \square^{\omega_i} \Phi_i
\right]
+																																
\I \xi \cdot \bar{\theta}^2 \;
\mathrm{tr}\left[ \Phi_1 \Phi_2 \Phi_3 \right]
+
\I \xi																															\cdot \theta^2 \;
\mathrm{tr}\left[ \Phi_1^\dagger \Phi_2^\dagger \Phi_3^\dagger \right]
\right\rbrace ~,
\end{split}
\label{eq:DSbetaDeformation_gen}
\end{equation}
where $\boldsymbol{\omega}$ is a shorthand notation for the deformation parameters $\omega_1$, $\omega_2$ and $\omega_3$.
It will be related to the model's spectral parameters, see \eqref{eq:gen_superPropagator_graphical} below.
The mass dimension of the chiral superfield gets deformed to $\left[\Phi_i\right] = [\Phi^\dagger_i ] = \frac{(D-2\mathcal{N})-2\omega_i}{2}\vert_{D=4,\mathcal{N}=1} = 1 - \omega_i$. For the interaction to be marginal, we require $3 - (\omega_1 +\omega_2 + \omega_3) = D - \mathcal{N}$, which is for $D=4$, $\mathcal{N}=1$ equivalent to the relation $\omega_1 +\omega_2 + \omega_3 = 0$. 
One recovers the original theory \eqref{eq:DSbetaDeformation} by setting all $\omega_i = 0$.
The kinetic terms of the component fields are
\begin{equation}
\begin{split}
S_{\mathrm{kin},\boldsymbol{\omega}} 
&=
\mathrm{N}\int \dd^4 x\; \dd^2\theta \dd^2 \bar{\theta} 
\sum_{i=1}^{3} 
\mathrm{tr} \left[ 
\Phi_i^\dagger \square^{\omega_i} \Phi_i
\right]\\
&=
\mathrm{N}\int \dd^4 x\; 
\sum_{i=1}^{3} 
\mathrm{tr} \left[ 
\phi_i^\dagger \square^{1+\omega_i} \phi_i -
\I \bar{\psi}_i \bar{\sigma}^\mu \square^{\omega_i} \partial_\mu \psi_i +
F_i^\dagger \square^{\omega_i} F_i
\right] ~,
\end{split}
\label{eq:DSbetaDeformation_gen_kin}
\end{equation}
and we will use their non-local kinetic operators to construct the superpropagator in section \ref{subsec:TheGeneralizedSuperpropagator} below.

Note that the $\boldsymbol{\omega}$-deformed action $S_{\boldsymbol{\omega}}$ in \eqref{eq:DSbetaDeformation_gen} is formally still supersymmetric.
Indeed, the kinetic part is still a D-term, expressed as a full superspace integral, such that the usual argument applies.
Defining $F(z)$ to be an arbitrary function of the $\mathcal{N} = 1$ supercoordinate $z = (x,\theta, \bar{\theta})$, we have  
\begin{equation}
0
\stackrel{\mathrm{!}}{=}
\delta_\varepsilon
\int \dd^4 x\; \dd^2\theta \dd^2 \bar{\theta} F(z)
=
\int \dd^4 x\; \dd^2\theta \dd^2 \bar{\theta}\; \delta_\varepsilon F(z)
\end{equation}
with 
\begin{equation}
\delta_\varepsilon F(z)
=
\left(
\varepsilon^\alpha Q_\alpha + \bar{\varepsilon}_{\dot{\alpha}} \bar{Q}^{\dot{\alpha}}
\right)
F(z)
=
\left(
\varepsilon^\alpha \partial_\alpha + \bar{\varepsilon}_{\dot{\alpha}} \bar{\partial}^{\dot{\alpha}}
\right)
F(z)
+ \mathrm{total~derivative}.
\end{equation}
Finally, the superspace integral over a Gra\ss mann derivative of $F(z)$ vanishes, 
\begin{equation}
\int \dd^4 x\; \dd^2\theta \dd^2 \bar{\theta}\; \partial_\alpha F(z)
=
0
=
\int \dd^4 x\; \dd^2\theta \dd^2 \bar{\theta}\; \bar{\partial}^{\dot{\alpha}} F(z) ~ ,
\end{equation}
since otherwise $F(z)$ would need to have a $\theta^3 \bar{\theta}^2$ (l.h.s.), respectively  $\theta^2 \bar{\theta}^3$ (r.h.s.), component.
The superpotential is unaltered and therefore still supersymmetric as well.
The supersymmetry of $S_{\boldsymbol{\omega}}$ is also consistent with a formal counting of the degrees of freedom, since the latter are in total unchanged due to the constraint $\omega_1 + \omega_2 + \omega_3 = 0$.
%\MK{
%I suppose that is the superspace approach to check for supersymmetry. However, by making the auxiliary field dynamical, the super multiplet grows by another dynamic field. Is this a proof for broken SUSY? There is a paper on non-local superfields and they require the auxiliary field to stay non-dynamical \cite{Kimura:2016irk}.
%}

\section{Stars, triangles, chains and super x-unity}

\subsection{The generalized superpropagator}
\label{subsec:TheGeneralizedSuperpropagator}
The generalized superfield propagator can be deduced from the individual generalized propagators of the component fields.
In fact, using \eqref{eq:Superfield_Components}, we find 
\begin{equation}
\begin{split}
\left\langle \Phi_i(z_1) \Phi_j^\dagger(z_2) \right\rangle 
=&~
\e^{
\I\theta_1 \sigma^\mu \bar{\theta}_1 \partial_{1,\mu} -
\I\theta_2 \sigma^\mu \bar{\theta}_2 \partial_{2,\mu}} \cdot \\
& \left[
\left\langle \phi_i(x_1) \phi_j^\dagger(x_2) \right\rangle 
+ 2 \theta_1^\alpha \bar{\theta}_2^{\dot{\alpha}}
\left\langle \psi_{i,\alpha}(x_1) \bar{\psi}_{j,\dot{\alpha}}(x_2) \right\rangle 
+ \theta_1^2 \bar{\theta}_2^2
\left\langle F_i(x_1) F_j^\dagger(x_2) \right\rangle
\right] ~.
\end{split}\label{eq:superPropagator_in_components}
\end{equation}
Here, $z_n = (x_n, \theta_n, \bar{\theta}_n)$ are again supercoordinates.
The generalized propagators of the components are derived from the inverse of the kinetic operators in the generalized action \eqref{eq:DSbetaDeformation_gen_kin} and we find
\begin{subequations}
\begin{align}
\left\langle \phi_i(x_1) \phi_j^\dagger(x_2) \right\rangle 
& =
\phantom{-\I} \delta_{ij}\; G_{1 + \omega_i}(x_{12}) ~, \\
\left\langle \psi_{i,\alpha}(x_1) \bar{\psi}_{j,\dot{\alpha}}(x_2) \right\rangle
& =
- \I \delta_{ij}\, \sigma^\mu_{\alpha \dot{\alpha}} \partial_{1,\mu}\; G_{1 + \omega_i}(x_{12}) ~, \\
\left\langle F_i(x_1) F_j^\dagger(x_2) \right\rangle
& =
\phantom{-\I}\delta_{ij} \square_1\; G_{1 + \omega_i}(x_{12}) ~.\label{eq:gen_Propagators_aux}
\end{align}\label{eq:gen_Propagators_components}%
\end{subequations}
Here we used the notation $x_{12}^\mu = x_1^\mu - x_2^\mu$ and
\begin{equation}
G_{u}(x) 
:=
\frac{1}{4\pi^2}
\int \dd^4 p \; \e^{-\I\, p \cdot x} \left[ -\frac{1}{p^2} \right]^u
=
- \left(-1 \right)^{u} 4^{1 - u}  \frac{\Gamma (2 - u)}{\Gamma (u)} \frac{1}{\left[ x^2 \right]^{2-u}} ~.
\end{equation}
For $u \rightarrow 0$, $G_{u}(x)$ becomes proportional to a delta function, see the representation \eqref{eq:DeltaDefi} below. 
This is in agreement with the expected delta function propagator for the non-dynamical auxiliary field \eqref{eq:gen_Propagators_aux} in the undeformed theory.
Finally, by plugging \eqref{eq:gen_Propagators_components} into \eqref{eq:superPropagator_in_components}, we obtain the generalized propagator of a chiral superfield
\begin{equation}
\begin{split}
\left\langle \Phi_i(z_1) \Phi_j^\dagger(z_2) \right\rangle
& =
\delta_{ij} (-4)^{- \omega_i} \frac{\Gamma (1 - \omega_i)}{\Gamma (1 + \omega_i)}
\e^{\I \left[ 
\theta_1 \sigma^\mu \bar{\theta}_1 +
\theta_2 \sigma^\mu \bar{\theta}_2 -
2 \theta_1 \sigma^\mu \bar{\theta}_2 \right]\partial_{1,\mu}}
\frac{1}{\left[ x_{12}^2 \right]^{1 - \omega_i}} \\
& =
\delta_{ij} (-4)^{- \omega_i} \frac{\Gamma (1 - \omega_i)}{\Gamma (1 + \omega_i)}
\frac{1}{\left[x_{1\bar{2}}^2 \right]^{1 - \omega_i}} ~.
\label{eq:gen_superPropagator}
\end{split}
\end{equation}
The exponential in \eqref{eq:gen_superPropagator} is a shift operator and produces the superconformally covariant interval 
$x_{1\bar{2}}^\mu := x_{12}^\mu+\I \left[ 
\theta_1 \sigma^\mu \bar{\theta}_1 +
\theta_2 \sigma^\mu \bar{\theta}_2 -
2 \theta_1 \sigma^\mu \bar{\theta}_2 \right]$.
Note that even though we are in Minkowski spacetime we did not explicitly include the $\I\varepsilon$ in the denominators of \eqref{eq:gen_superPropagator} for conciseness of notation. 

Graphically the generalized superpropagator is represented by 
\begin{equation}
\left\langle \Phi_i(z_1) \Phi_i^\dagger(z_2) \right\rangle
=
(-4)^{- \omega_i} a(1 + \omega_i) \cdot
\adjincludegraphics[valign=c,scale=1]{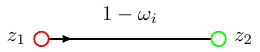} ,
\label{eq:gen_superPropagator_graphical}
\end{equation}
where the little arrow indicates the chiral end and $a(u)$ is a normalization defined below in \eqref{eq:RfactorAfactor}.
Observe that by tuning all spectral parameters to $\omega_i = 0$, we recover our theory of interest \eqref{eq:DSbetaDeformation} and the generalized superpropagator \eqref{eq:gen_superPropagator} reduces to the ordinary superpropagator $\frac{1}{x_{1\bar{2}}^2}$ derived in \cite{Wess:1992cp}.
The prefactor in \eqref{eq:gen_superPropagator_graphical} is just the proper normalization of the super Feynman diagrams under investigation.
Accordingly, we will focus in the following on the weight function
\begin{equation}
\frac{1}{\left[x_{1\bar{2}}^2 \right]^{u}}
=
\adjincludegraphics[valign=c,scale=1]{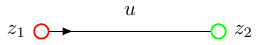} ~,
\label{eq:gen_superWeight_graphical}
\end{equation}
which happens to be the supersymmetric generalization of eq.\ (3.1) in \cite{Kade:2023xet}, and  where $u\in \mathbbm{C}$  is the spectral parameter needed to exploit the model's integrability.

\subsection{Superconformal star integral}
\label{subsec:SuperconfStarIntegral}
The main new tool in this paper is Osborn's star integral \cite{Osborn:1998qu}.
This is a superspace integral over a point in superspace, which is connected to $n$ fixed superspace coordinates by $n$ superspace propagators \eqref{eq:gen_superWeight_graphical}.
This $n$-point function is subject to the constraints of superconformal symmetry.
In fact, for $n=3,4$ explicit forms have been determined in \cite{Dolan:2000uw}.
We are particularly interested in the case $n=3$, where the result reads
\begin{equation}
\begin{split}
& \I\int \dd^4x_0\; \dd^2\theta_0\, \dd^2 \bar{\theta}_0\; \delta^{(2)}(\theta_0)\;
\frac{1}{\left[x_{1\bar{0}}^2 \right]^{u_1}}
\frac{1}{\left[x_{2\bar{0}}^2 \right]^{u_2}}
\frac{1}{\left[x_{3\bar{0}}^2 \right]^{u_3}}\\
& \stackrel{u_1+u_2+u_3=3}{=}
-4\,
r (u_1, u_2, u_3)\;
\frac
{
\left(\theta_{12}\theta_{13}\right) x_{23,+}^2 +
\left(\theta_{23}\theta_{21}\right) x_{31,+}^2 +
\left(\theta_{31}\theta_{32}\right) x_{12,+}^2 
}
{
[ x_{12,+}^2 ]^{2-u_3}
[ x_{23,+}^2 ]^{2-u_1}
[ x_{31,+}^2 ]^{2-u_2}
} ~.
\end{split}
\label{eq:Osborn_original}
\end{equation}
Here, we have used the abbreviations $x_{ij,+}^\mu := x_{i,+}^\mu - x_{j,+}^\mu$ as well as\footnote{These correspond to $r_0(u_1,u_2,u_3)$ and $a_0 (u) $ in \cite{Kade:2023xet}.}
\begin{equation}
r(u_1,u_2,u_3):= \pi^2 a (u_1) a (u_2) a (u_3)
~~\mathrm{with}~~
a (u) := \frac{\Gamma (2 - u )}{\Gamma ( u )} ~.
\label{eq:RfactorAfactor}
\end{equation}
Note the similarity of \eqref{eq:Osborn_original} with a star-triangle relation (STR), see e.g. \cite{Chicherin:2012yn,DEramo:1971hnd,Symanzik:1972wj}.
Generally, a STR is a relation between a subgraph in the shape of a three-spiked star and a triangle, modulo some factor, which depends on the model under investigation.
This graphical move is sufficient to construct an R-matrix, which in turn is the building block for the construction of commuting transfer matrices \cite{Bazhanov:2011mz}, one of the hallmarks of integrability.
However, the r.h.s. of \eqref{eq:Osborn_original} is not quite of the form of a triangle built from superpropagators of the type \eqref{eq:gen_superPropagator_graphical}, due to the non-factorizing numerator.

Unfortunately, we have not yet been able to derive a suitable Yang-Baxter equation (YBE) along with a manifestly commuting transfer matrix based on \eqref{eq:Osborn_original}.
We are however confident that these exist, postponing their rigorous derivation to later work.
In this context, we would like to make the important remark that a proper STR is in general only a sufficient but not necessary condition for an YBE.
Remarkably, despite this shortcoming, in the below we will provide exciting evidence for the model's integrability by demonstrating that Zamolodchikov's method of inversions \cite{Zamolodchikov:1980mb}, see also \cite{Kade:2023xet}, may nevertheless be successfully applied.

\subsection{Super chain relations}
\label{subsec:SuperChainRelations}
Similar to the bosonic STR, we can derive superspace chain relations for the convolution of two superspace propagators \eqref{eq:gen_superPropagator_graphical}. 
To this end, we could take the bosonic coordinate of one external point in \eqref{eq:Osborn_original} to infinity and compare the proportionality constants.
Alternatively, we can show by direct integration, see appendix \ref{subsec:SuperChainRule}, the chain relation
\begin{subequations}
\begin{align}
\left[
\I \int \dd^4x_0\, \dd^2\bar{\theta}_0
\frac{1}{\left[x_{1\bar{0}}^2 \right]^{u_1}}
\frac{1}{\left[x_{2\bar{0}}^2 \right]^{u_2}}
\right]_{\substack{\theta_0 = 0 \\ \bar{\theta}_{1,2}=0}}
=
 - 4\, r ( 3 - u_1 - u_2 , u_1 , u_2 )\;
\frac{\theta_{12}^2}{\left[x_{12}^2 \right]^{u_1+u_2 -1}} ~, \\
\adjincludegraphics[valign=c,scale=0.7]{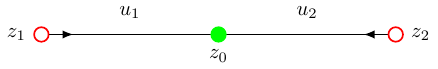}
= 
 - 4\, r ( 3 - u_1 - u_2 , u_1 , u_2 )\;
\adjincludegraphics[valign=c,scale=0.7]{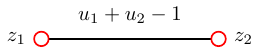} ~,
\end{align}
\label{eq:ChainRelAntiChiral}%
\end{subequations}
and its chiral counterpart
\begin{subequations}
\begin{align}
\left[
\I \int \dd^4x_0\, \dd^2\theta_0
\frac{1}{\left[x_{\bar{1}0}^2 \right]^{u_1}}
\frac{1}{\left[x_{\bar{2}0}^2 \right]^{u_2}}
\right]_{\substack{\bar{\theta}_0 = 0 \\ \theta_{1,2}=0}}
=
 - 4\, r ( 3 - u_1 - u_2 , u_1 , u_2 )\;
\frac{\bar{\theta}_{12}^2}{\left[x_{12}^2 \right]^{u_1+u_2 -1}} ~, \\
\adjincludegraphics[valign=c,scale=0.7]{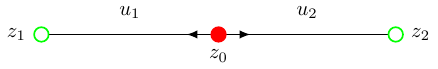}
= 
 - 4\, r ( 3 - u_1 - u_2 , u_1 , u_2 )\;
\adjincludegraphics[valign=c,scale=0.7]{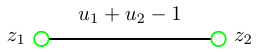} ~.
\end{align}
\label{eq:ChainRelChiral}%
\end{subequations}
In the last two equations, we introduced new graphical representations for the following formal two-point functions in superspace
\begin{equation}
\frac{\theta_{12}^2}{\left[x_{12}^2 \right]^{u}} =
\adjincludegraphics[valign=c,scale=0.7]{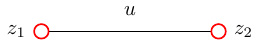} ~,
\hspace{1cm}
\frac{\bar{\theta}_{12}^2}{\left[x_{12}^2 \right]^{u}} =
\adjincludegraphics[valign=c,scale=0.7]{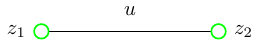} ~,
\label{eq:AuxTwoPointFctns}
\end{equation}
which act like chiral and anti-chiral delta functions on the fermionic subspace, respectively.
On the bosonic part of superspace, the functions \eqref{eq:AuxTwoPointFctns} are bosonic generalized propagators. This means that they are subject to the usual bosonic delta function prescription when their spectral parameter approaches $D/2 = 2$ in conjunction with being multiplied by $a (\varepsilon)$,
\begin{equation}
\delta^{(4)}\left( x_{12} \right) 
~=~
\lim_{\varepsilon \rightarrow 0} ~ \pi^{-2} a (\varepsilon)
\cdot \frac{1}{\left[ x_{12}^2 \right]^{2 - \varepsilon}} ~.
\label{eq:DeltaDefi}
\end{equation}
We are therefore able to express the unity kernel for chiral/anti-chiral superspace integrations through the convolution of two generalized superspace propagators as
\begin{subequations}
\begin{align}
\lim_{\varepsilon \rightarrow 0}
\adjincludegraphics[valign=c,scale=0.7]{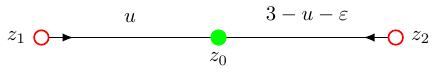}
&= 
 - 4\,\pi^4\cdot a ( u )\, a ( 3 - u )
\adjincludegraphics[valign=c,scale=0.7]{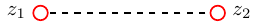} ~,
\label{eq:ResolutionOfUnity_antichiral}\\
\lim_{\varepsilon \rightarrow 0}
\adjincludegraphics[valign=c,scale=0.7]{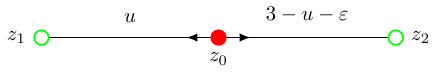}
&= 
 - 4\,\pi^4\cdot a( u )\, a ( 3 - u )
\adjincludegraphics[valign=c,scale=0.7]{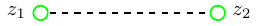} ~. \label{eq:ResolutionOfUnity_chiral}
\end{align}
\label{eq:ResolutionOfUnity}%
\end{subequations}
Here the chiral and anti-chiral delta functions are represented graphically as 
\begin{equation}
\delta^{(2)}\left(\theta_{12}\right) \delta^{(4)}\left( x_{12} \right) =
\adjincludegraphics[valign=c,scale=0.7]{pictures/basics/deltadist/chiral_delta/DeltaDist.pdf} ~,
\hspace{0.7cm}
\delta^{(2)}\left(\bar{\theta}_{12}\right) \delta^{(4)}\left( x_{12} \right) =
\adjincludegraphics[valign=c,scale=0.7]{pictures/basics/deltadist/antichiral_delta/DeltaDist.pdf} ~.
\label{eq:SuperspaceDeltaDefi}
\end{equation}

%In order to do so, we first derive an auxiliary relation, called x-unity. All the ingredients therefore can be derived from Osborn's formula \cite{Osborn:1998qu} for an anti-chiral star of superfield propagators
%\begin{equation}
%\begin{split}
%\adjincludegraphics[valign=c,scale=0.6]{pictures/basics/str/antichiral_superfield/STR_LHS.pdf} 
%\sim
%\left[
%\int \dd^4x_0\; \dd^2\theta_0\, \dd^2 \bar{\theta}_0\; \delta^{(2)}(\theta_0)
%\frac{\I}{\left[x_{1\bar{0}}^2 \right]^{u_1}}
%\frac{\I}{\left[x_{2\bar{0}}^2 \right]^{u_2}}
%\frac{\I}{\left[x_{3\bar{0}}^2 \right]^{u_3}}
%\right]_{\bar{\theta}_{1,2,3}=0}\\
%\stackrel{u_1+u_2+u_3=3}{=}
%4\pi^2 
%\left[
%\prod_{i=1}^3
%\frac{\Gamma (2-u_i)}{\Gamma (u_i)}
%\right]
%\frac
%{
%\left(\theta_{12}\theta_{13}\right) x_{23}^2 +
%\left(\theta_{23}\theta_{21}\right) x_{31}^2 +
%\left(\theta_{31}\theta_{32}\right) x_{12}^2 
%}
%{
%\left[ x_{12}^2 \right]^{2-u_3}
%\left[ x_{23}^2 \right]^{2-u_1}
%\left[ x_{31}^2 \right]^{2-u_2}
%}.
%\end{split}
%\label{eq:Osborn}
%\end{equation}
%A similar relation holds for the red chiral star super-diagram.

\subsection{Super x-unity relation}
Before forging, as in \cite{Kade:2023xet}, our main tool, the super x-unity relation, we make two important observations regarding the star integral \eqref{eq:Osborn_original}:
\begin{itemize}
\item 
If we take one superpropagator weight to zero in \eqref{eq:Osborn_original}, we obtain \eqref{eq:ResolutionOfUnity} with one external point detached from the rest of the diagram. 
In detail, taking the limit $u_1 = \varepsilon \rightarrow 0$ of \eqref{eq:Osborn_original} yields
\begin{equation}
\lim_{\varepsilon\rightarrow 0}
\adjincludegraphics[valign=c,scale=0.6]{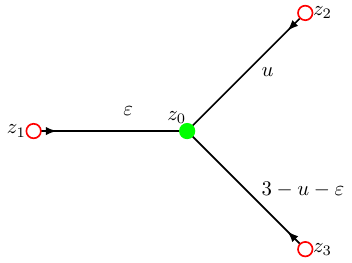} 
=
- 4\,\pi^4
a(u)\, a(3 - u) ~\cdot
\adjincludegraphics[valign=c,scale=0.6]{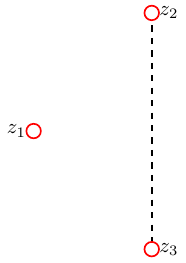} ~.
\label{eq:3ptFctnWithOneVanishingParameter}
\end{equation}

\item
Another helpful relation is obtained when integrating one external point of \eqref{eq:Osborn_original} over the chiral subspace of superspace, namely
\begin{equation}
\adjincludegraphics[valign=c,scale=0.6]{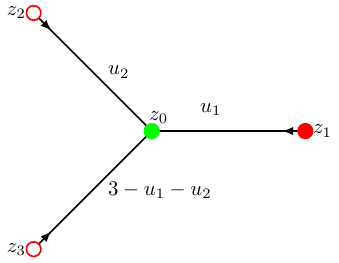} 
=
- 4\,\pi^4
a(u_1)\, a(3 - u_1)
\adjincludegraphics[valign=c,scale=0.6]{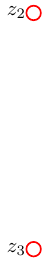} ~.
\label{eq:3ptFctnOnePointIntegrated}
\end{equation}
We observe that the whole expression reduces to a $u_2$-independent factor.
\end{itemize}

Having established the chain relations in section \ref{subsec:SuperChainRelations} and the relations \eqref{eq:3ptFctnWithOneVanishingParameter} and \eqref{eq:3ptFctnOnePointIntegrated}, we are now able to use them to derive our main auxiliary relation, a supersymmetric generalization of the bosonic x-unity relation \cite{Kade:2023xet}.
Details are given in appendix \ref{subsec:SuperXunity} in equation \eqref{eq:SuperXUnity_derivation}.
As in the bosonic case, the name indicates that the relation reduces an x-shaped supergraph to a single super-delta function.
We can perform the derivation also for the chirality-inverted x-shaped supergraph, therefore we get two equations 
\begin{subequations}
\begin{align}
\adjincludegraphics[valign=c,scale=0.6]{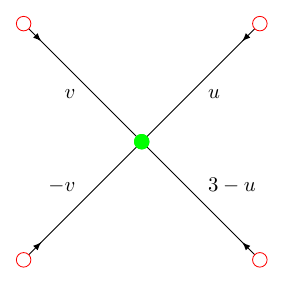} 
=
- 4\,\pi^4
a(u)\, a(3-u) \hspace{0.5cm}
\adjincludegraphics[valign=c,scale=0.6]{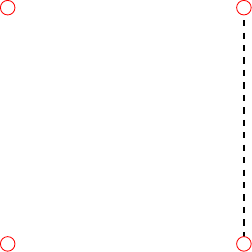} ~,\label{eq:SuperXUnity_green}\\
\adjincludegraphics[valign=c,scale=0.6]{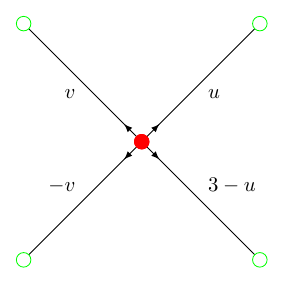} 
=
- 4\,\pi^4
a(u)\, a(3-u) \hspace{0.5cm}
\adjincludegraphics[valign=c,scale=0.6]{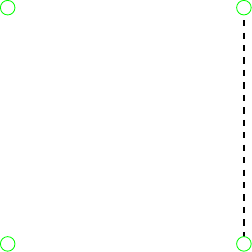} ~.\label{eq:SuperXUnity_red}
\end{align}\label{eq:SuperXUnity}%
\end{subequations}
These relations are the most crucial tool in the computation of the model's critical coupling, to which we turn next.

\section{Vacuum graphs in the thermodynamic limit}

\begin{figure}[h]
\includegraphics[scale=0.5]{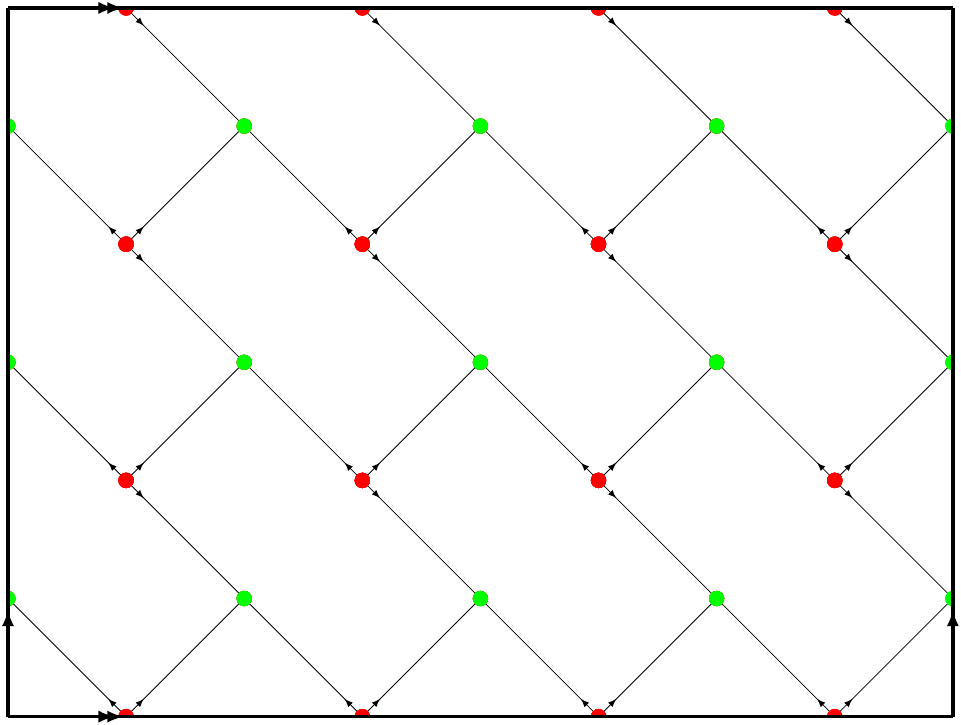} \centering
\caption{The toroidal super vacuum graph $Z_{3,4}$ is shown as an example for a contribution to the double-scaled $\beta$-deformed SYM's free energy.
% It contains three rows, each consisting of four graph building kernels \eqref{eq:GraphBuilderBW}. 
The faces of the graph are hexagons. In the limit $N\rightarrow\infty$, the leading order results in toroidal diagrams. This is implemented into the figure: 
%the thick lines bounding the figure should be identified according to the number of arrows on them: 
top and bottom lines are identified, and so are the left and right boundary lines.}
\label{fig:betaDef_vacuumgraph}
\end{figure}

The vacuum supergraphs of the double-scaled $\beta$-deformation \eqref{eq:DSbetaDeformation} in the planar limit are shown in fig.\ \ref{fig:betaDef_vacuumgraph}. 
The regular brick wall (honeycomb) pattern is due to the highly constraining Feynman rules and notably the interaction vertices \eqref{eq:betaDef_FeynmanRules_vertices} in the action \eqref{eq:DSbetaDeformation}. 
Diagrammatically, the generalized double-scaled $\beta$-deformation \eqref{eq:DSbetaDeformation_gen} reproduces the same graphs as in fig.\ \ref{fig:betaDef_vacuumgraph}, because the superpotential is the same as in the undeformed theory \eqref{eq:DSbetaDeformation}.
However, the propagator weights differ by their exponent and one has to multiply with the prefactor appearing in \eqref{eq:gen_superPropagator_graphical}.

An important comment is in place: vacuum diagrams in field theory are generally proportional to the spacetime volume and thus the overall free energy is usually infrared divergent. 
Since we are in any case only interested in the free energy density, this divergence is best fixed by leaving one of the points in the vacuum graphs unintegrated, which fixes the zero mode.
Similarly, superspace vacuum graphs are proportional to the ill-defined $\int \dd^4 x\cdot\int \dd^2 \theta\, \dd^2 \bar{\theta} = \infty \cdot 0$.
(The zero stemming from the fermionic integration is just a manifestation of the well known statement that the vacuum energy of supersymmetric field theories is zero.)
In our supersymmetric model at hand, we proceed in exactly the same way and leave one of the superspace points in the vacuum supergraphs unintegrated, in order to obtain a well-defined density. 

\begin{figure}[h]
\includegraphics[scale=0.7]{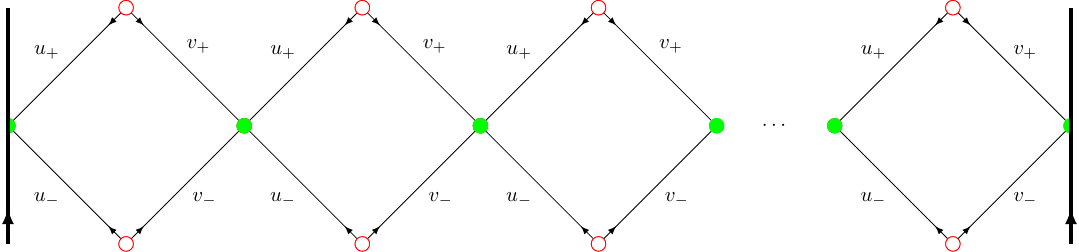}  \centering
\caption{The generalized row matrix $T_N(\mathbf{u})$ contains anti-chiral internal vertices (green, filled). Its external points are chiral, un-integrated vertices (red, un-filled). The row matrix depends on four spectral parameters, collectively denoted as $\mathbf{u}=\left(\begin{smallmatrix}u_+ & v_+ \\ u_- & v_- \end{smallmatrix}\right)$. Multiple copies of the row matrix can be stacked on top of each other to build up a generalized super Feynman graph according to \eqref{eq:gen_vacuumdiagram}. The ``matrix product'' thereby consists of $N$ integrals over the chiral part of superspace.}
\label{fig:betaDef_gen_rowmatrix}
\end{figure}

Starting from generalized propagators \eqref{eq:gen_superWeight_graphical}, we can identify\footnote{ 
Note, that we could have chosen another row matrix kernel with the chiralities interchanged, i.\ e.\ with anti-chiral (green) external vertices and chiral (red) internal ones.
}
a generalized row-matrix $T_N(\mathbf{u})$, see fig.\ \ref{fig:betaDef_gen_rowmatrix}, building up the vacuum diagrams in fig.\ \ref{fig:betaDef_vacuumgraph} after e.\ g.\ fixing the parameters to $\mathbf{u}:=\left(\begin{smallmatrix}u_+ & v_+ \\ u_- & v_- \end{smallmatrix}\right)=\left(\begin{smallmatrix}0 & 1 \\ 1 & 1 \end{smallmatrix}\right)$.
Formally, we may write a generalized $M\x N$ toroidal vacuum supergraph as 
\begin{equation}
Z_{MN}(\mathbf{u}) = \mathrm{tr}\left[ T_N(\mathbf{u})^M \right],
\label{eq:gen_vacuumdiagram}
\end{equation}
which graphically represents $M$ generalized row matrices of length $N$ stacked on top of each other and identified periodically by the trace. 

For the more general supergraphs of theory \eqref{eq:DSbetaDeformation_gen} we have to tune the parameters to e.\ g.\ $\mathbf{u}=\boldsymbol{\omega}:=\left(\begin{smallmatrix}0 & 1 - \omega_1 \\ 1 - \omega_3 & 1 - \omega_2 \end{smallmatrix}\right)$ and multiply the corresponding factors from the propagators \eqref{eq:gen_superPropagator_graphical} to the row matrix.
We find 
\begin{equation}
\mathbbm{T}_N (\mathbf{u}) 
=
\left[
a (2 - u_-)\, a (2 - v_+)\, a(2 - v_-)
\right]^N
T_N(\mathbf{u})
\label{eq:gen_RowMatrix_omega}
\end{equation}
to be a suitable row matrix to build the vacuum diagrams\footnote{Remember that $\omega_1 + \omega_2 + \omega_3 = 0$, which annihilates the product of the $(-4)^{-\omega_i}$ for each flavor from \eqref{eq:gen_superPropagator_graphical}.
}.
We denote the generalized $M\x N$ toroidal vacuum supergraphs built by the enhanced row matrix by
\begin{equation}
\mathbbm{Z}_{MN}(\mathbf{u}) = \mathrm{tr}\left[ \mathbbm{T}_N(\mathbf{u})^M \right] ~.
\label{eq:gen_vacuumdiagram_enhanced}
\end{equation}
They are the super vacuum diagrams of, first, the generalized double-scaled $\beta$-deformed SYM theory \eqref{eq:DSbetaDeformation_gen} when specifying the spectral parameters to the value $\mathbf{u}\rightarrow \boldsymbol{\omega}$ and, second, the double-scaled $\beta$-deformed SYM theory \eqref{eq:DSbetaDeformation} when $\boldsymbol{\omega} \rightarrow \left(\begin{smallmatrix}0 & 1 \\ 1 & 1 \end{smallmatrix}\right)$ or equivalently all $\omega_i \rightarrow 0$. 
By way of example, the resulting graph for $M=3$ and $N=4$, called $Z_{34}\left(\begin{smallmatrix}0 & 1 \\ 1 & 1 \end{smallmatrix}\right) = \mathbbm{Z}_{34}\left(\begin{smallmatrix}0 & 1 \\ 1 & 1 \end{smallmatrix}\right)$, is shown in fig. \ref{fig:betaDef_vacuumgraph}.

Our goal is to calculate the critical coupling $\xi_\mathrm{cr}$. It is defined as the radius of convergence of the expansion of the free energy of \eqref{eq:DSbetaDeformation_gen} (including the special case \eqref{eq:DSbetaDeformation}), which is 
\begin{equation}
Z_{\boldsymbol{\omega}} ~=~ 
\sum_{M,N=1}^\infty \mathbbm{Z}_{MN}\left(\boldsymbol{\omega}\right) \cdot 
\left( - \xi \right)^{2MN} ~ .
\label{eq:gen_freeEnergySeries}
\end{equation}
The critical coupling is then $\xi_\mathrm{cr}= \left[ \mathbbm{K}\left(\boldsymbol{\omega}\right) \right]^{-1/2}$, where we evaluate the generalized vacuum diagrams in the thermodynamic limit 
\begin{equation}
\mathbbm{K}(\mathbf{u}) ~:=~
\lim_{M,N\rightarrow \infty} \vert \mathbbm{Z}_{MN}(\mathbf{u})\vert^{\frac{1}{MN}}
\label{eq:gen_freeEnergy}
\end{equation}
at $\mathbf{u}=\boldsymbol{\omega}$ and eventually at $\omega_i \rightarrow 0$. 

At this point, we comment on the free energy of non-supersymmetric double-scaled $\gamma$-deformations, which is the theory where every interaction term in the square brackets of \eqref{eq:DS_components} multiplies a different coupling. 
Due to the broken supersymmetry, we can no longer recast the component action into a superspace action.
However, collecting component Feynman graphs into formal supergraphs is still a sensible thing to do because the vacuum supergraphs $\mathbbm{Z}_{MN}\left(\boldsymbol{\omega}\right)$ do not depend on the coupling.
Still, one has to consider now that different components of the same supergraph will enter the free energy \eqref{eq:gen_freeEnergySeries} with different couplings, and in order to determine them, one has to decompose the supergraph into its components.

We will calculate the limit \eqref{eq:gen_freeEnergy} by the method of inversion relations.
As a first step, this requires finding the inverse of the row matrix $T_N(\mathbf{u})$ and determining the limit $K (\mathbf{u}):=\lim_{M,N\rightarrow \infty} \vert Z_{MN}(\mathbf{u})\vert^{\frac{1}{MN}}$.
Subsequently, we make the connection to \eqref{eq:gen_freeEnergy} by reinstating the additional factors in \eqref{eq:gen_RowMatrix_omega} via the relation
\begin{equation}
\mathbbm{K}(\mathbf{u}) 
~=~
\left[
a (2 - u_-)\, a(2 - v_+)\, a(2 - v_-)
\right]
K(\mathbf{u}) ~.
\label{eq:mathbbmK_to_K}
\end{equation}

\subsection{Inversion relations}
The super x-unity relation allows us to find four different forms of the inverse of the row matrix $T_N(\mathbf{u})$, by the same moves explained in section 5 of \cite{Kade:2023xet}. We find
\begin{equation}
T_N(\mathbf{u})
\circ
T_N(\mathbf{u}_{\mathrm{inv}})
=
F_N \cdot \mathbbm{1}_N
\label{eq:RowMatrixInversion}
\end{equation}
to hold for 
\begin{subequations}
\begin{align}
\mathbf{u}_\mathrm{inv}&=\left(\begin{smallmatrix} -u_- & 3-v_- \\ 3-u_+ & -v_+ \end{smallmatrix}\right)~~\mathrm{and}~~F_N=\left[ 16\pi^8\; a(u_+)\, a(3-u_+)\, a(v_-)\, a(3-v_-) \right]^N ~, \\
\mathbf{u}_\mathrm{inv}&=\left(\begin{smallmatrix} -u_- & 3-v_- \\ -u_+ & 3-v_+ \end{smallmatrix}\right)~~\mathrm{and}~~F_N=\left[ 16\pi^8\; a(v_+)\, a(3-v_+)\, a(v_-)\, a(3-v_-) \right]^N ~, \\
\mathbf{u}_\mathrm{inv}&=\left(\begin{smallmatrix} 3-u_- & -v_- \\ 3-u_+ & -v_+ \end{smallmatrix}\right)~~\mathrm{and}~~F_N=\left[ 16\pi^8\; a(u_+)\, a(3-u_+)\, a(u_-)\, a(3-u_-) \right]^N ~,\\
\mathbf{u}_\mathrm{inv}&=\left(\begin{smallmatrix} 3-u_- & -v_- \\ -u_+ & 3-v_+ \end{smallmatrix}\right)~~\mathrm{and}~~F_N=\left[ 16\pi^8\; a(u_-)\, a(3-u_-)\, a(v_+)\, a(3-v_+) \right]^N . \label{eq:RowMatrixInversion_4}
\end{align}\label{eq:RowMatrixInversion_1234}
\end{subequations}
For example, \eqref{eq:RowMatrixInversion_4} can be obtained from the super x-unity relation \eqref{eq:SuperXUnity} by the steps
\begin{subequations}
\begin{align}
T&_{N} 
\left(\begin{smallmatrix}u_+ & v_+ \\ u_- & v_- \end{smallmatrix}\right)
\circ
T_{N}
\left(\begin{smallmatrix} 3 - u_- & -v_- \\ -u_+ & 3 - v_+ \end{smallmatrix}\right)\\
~&\stackrel{\phantom{\eqref{eq:SuperXUnity_green}}}{=}~ 
\adjincludegraphics[valign=c,scale=0.5]{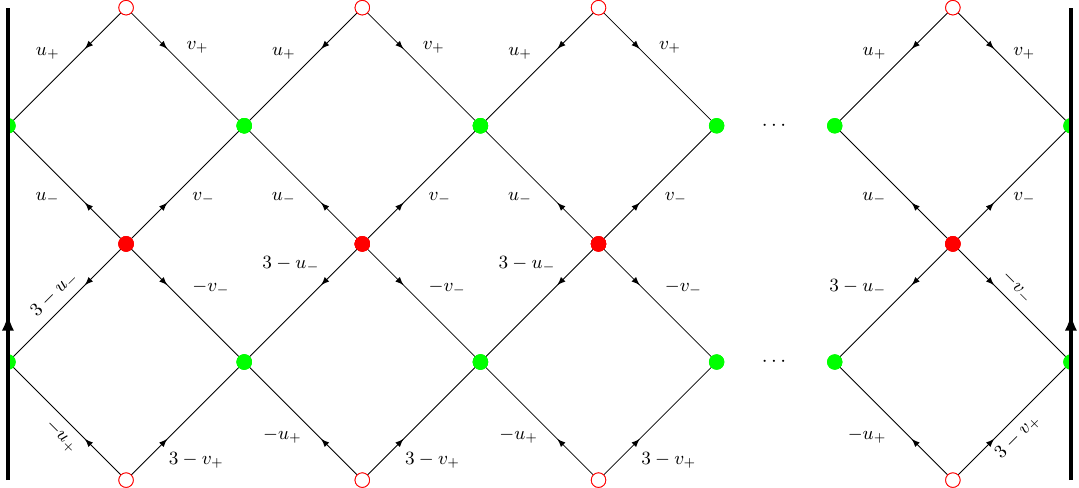}
\label{eq:TransferMatrixAnnihilation1}\\
%%%%%%%%%%
~&\stackrel{\eqref{eq:SuperXUnity_red}}{=}~
\adjincludegraphics[valign=c,scale=0.5]{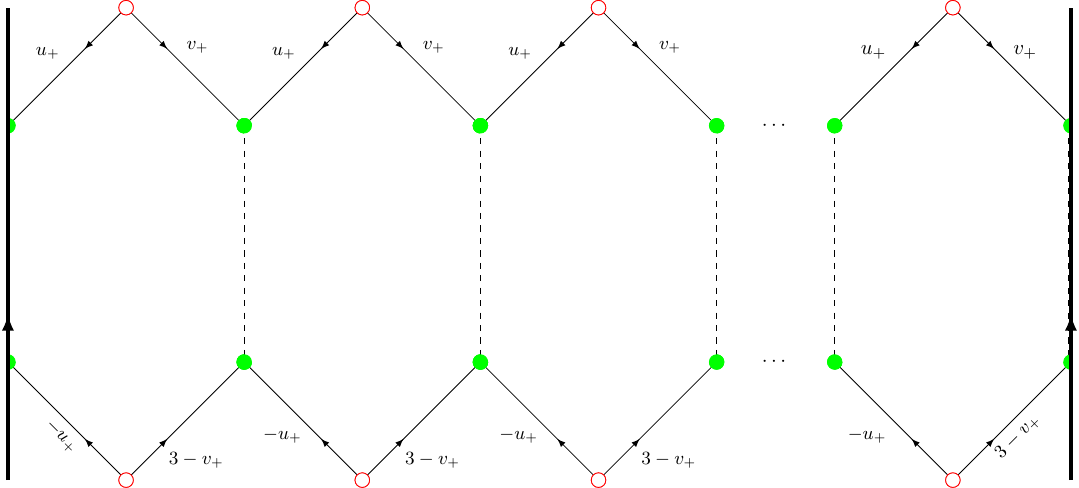} 
\cdot \nonumber \\
%%%%%
& \hspace{1.5cm} \cdot 
\left[
- 4 \pi^4 a (u_-) a (3 - u_-)
\right]^N 
\\
%%%%%%%%%%
~&\stackrel{\phantom{\eqref{eq:SuperXUnity_green}}}{=}~
\adjincludegraphics[valign=c,scale=0.5]{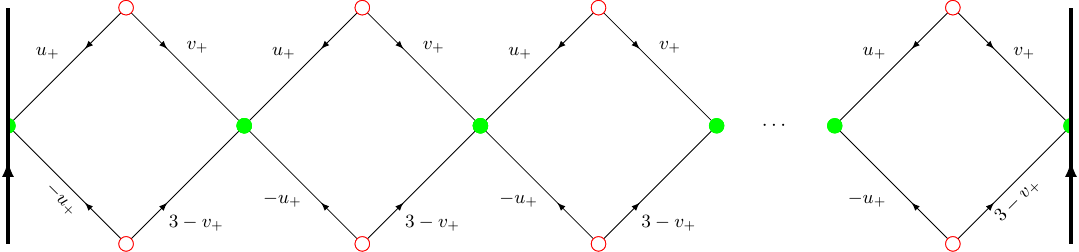}
\cdot \nonumber \\
%%%%%
& \hspace{1.5cm} \cdot 
\left[
- 4 \pi^4 a (u_-) a (3 - u_-)
\right]^N 
\\
%%%%%%%%%%
~&\stackrel{\eqref{eq:SuperXUnity_green}}{=}~
\adjincludegraphics[valign=c,scale=0.5]{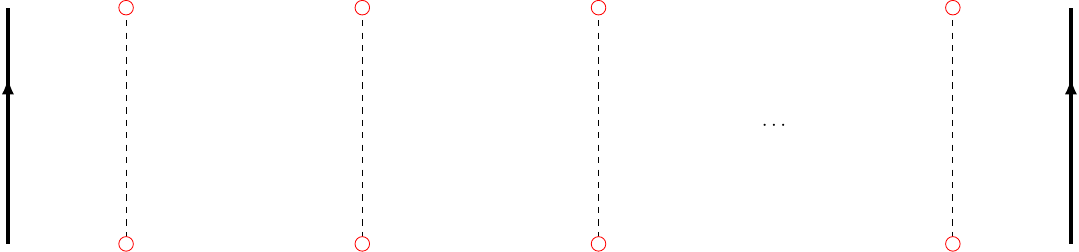} 
\cdot \nonumber \\
%%%%%
& \hspace{1.5cm} \cdot
\left[
16 \pi^8\; a (u_-) a (3 - u_-)\;
 a (v_+) a (3 - v_+)
\right]^N
\\
%%%%%%%%%%
~&\stackrel{\phantom{\eqref{eq:SuperXUnity_green}}}{=}
\left[
16 \pi^8\; a (u_-) a (3 - u_-)\;
 a (v_+) a (3 - v_+)
\right]^N \cdot \mathbbm{1}_N
\label{eq:TransferMatrixAnnihilationEnd} ~ ,
\end{align}\label{eq:TransferMatrixAnnihilation}%
\end{subequations}
and the other equations of \eqref{eq:RowMatrixInversion_1234} can be obtained by using variations of \eqref{eq:SuperXUnity} with left- and right external points interchanged.

Projecting \eqref{eq:gen_vacuumdiagram} on the eigenvector corresponding to the maximal eigenvalue dominating the thermodynamic limit \cite{Kade:2023xet},
\begin{equation}
K (\mathbf{u})
=
\lim_{M,N\rightarrow \infty} 
\vert \mathrm{tr}\left[ T_N(\mathbf{u})^M \right] \vert^{\frac{1}{MN}}
=
\lim_{N\rightarrow \infty} 
\vert \Lambda_{\mathrm{max},N}(\mathbf{u}) \vert^{\frac{1}{N}} ~,
\end{equation}
we can turn \eqref{eq:RowMatrixInversion} into four functional relations for $K(\mathbf{u})$. 
Based on the observation that $K(\mathbf{u})$ corresponds to a rhombus of four superpropagators, according to \eqref{eq:gen_freeEnergy} and \eqref{eq:mathbbmK_to_K}, we make the ansatz $K(\mathbf{u}) = \kappa_1(u_+)\kappa_2(u_-)\kappa_3(v_+)\kappa_4(v_-)$.
We find $\kappa_i(u) \equiv \kappa(u)$ for all $i$, which has to satisfy
\begin{equation}
\kappa(u)\kappa(-u) = 1
~~~
\mathrm{and}
~~~
\kappa(u)\kappa(3-u) = 4\pi^4\; a(u)\, a(3-u) = 4\pi^4 \frac{\Gamma (2-u) \Gamma (u-1)}{\Gamma (u) \Gamma(3-u)} ~.
\label{eq:InversionRelations_kappa}
\end{equation}
We can construct a solution using the method explained in \cite{Kade:2023xet}, which consists of plugging the two functional relations \eqref{eq:InversionRelations_kappa} iteratively into each other and requiring the solution to have no poles in the physical interval $\left[ 0, 2\right)$ corresponding to the maximal eigenvalue \cite{StroganovInvRelLatticeModels}.
We find
\begin{equation}
\kappa (u) 
=
12^{\frac{u}{3}} \pi ^{\frac{4 u}{3}} 
\frac
{\Gamma \left(\frac{u+1}{3}\right) \Gamma (2 -u) }
{\Gamma \left(\frac{1}{3}\right)}
\prod _{k=1}^{\infty } 
\frac
{\Gamma (3 k - u + 2 ) \Gamma ( 3 k + u ) \Gamma (3 k - 2 )}
{\Gamma (3 k + u - 2 ) \Gamma ( 3 k - u ) \Gamma (3 k + 2 )} ~.
\end{equation}
With the help of the functional relation of the $\Gamma$-function, we observe that the infinite product collapses to the expression
\begin{equation}
\kappa (u) 
=
2^{\frac{2 u}{3}} 3^{\frac{4 u}{3}-2} \pi ^{\frac{4 u}{3}}
\frac{\Gamma (2-u) \Gamma \left(\frac{u}{3}\right) \Gamma \left(\frac{u+1}{3}\right)}{\Gamma \left(1-\frac{u}{3}\right) \Gamma \left(\frac{4}{3}-\frac{u}{3}\right) \Gamma (u)}
\end{equation}
with the special values $\kappa (0) = 1$ and $\kappa (1) = \left(\frac{2\pi^2}{3} \right)^{2/3} \Gamma (\frac{1}{3})$.
The function $\kappa (u)$ has a characteristic pole at the value of the crossing parameter $u = 2$, which is the value for which the bosonic part of the superpropagator becomes almost a delta function, up to the damping factor $a(\varepsilon)$, c.\,f.\ \eqref{eq:DeltaDefi}.
Finally, we find the critical coupling
\begin{equation}
\xi_\mathrm{cr}
=
\left[ \mathbbm{K}\left(\begin{smallmatrix}0 & 1 \\ 1 & 1 \end{smallmatrix}\right) \right]^{-1/2}
= 
\kappa\left( 1 \right)^{-3/2}
=
\frac{3}{2 \pi ^2 \;\Gamma (\frac{1}{3})^{3/2}} ~
\label{eq:CritCoup_DS}
\end{equation}
for the double-scaled $\beta$-deformation of $\mathcal{N}=4$ SYM and 
\begin{equation}
\xi_\mathrm{cr}
=
\left[ \mathbbm{K}\left(\begin{smallmatrix}0 & 1 - \omega_1 \\ 1 - \omega_3 & 1 - \omega_2 \end{smallmatrix}\right) \right]^{-1/2}
= 
\left[
\prod_{i=1}^3
a( 1 + \omega_i )\,
\kappa\left( 1 - \omega_i \right)
\right]^{-1/2} 
\label{eq:CritCoup_DS_gen}
\end{equation}
for the deformed theory \eqref{eq:DSbetaDeformation_gen}.

\section{Conclusions and Outlook}
We summarize our results.
The superspace action of the double-scaled $\mathcal{N}=1$ $\beta$-deformation was obtained in \eqref{eq:DSbetaDeformation}.
It was found that the supergraphs admit a very regular brick wall structure, which suggest that supergraphs are the superior packaging of component Feynman diagrams in order to highlight integrable structures therein.
We introduced the generalized propagator of chiral superfields containing the spectral parameter and proposed a new action \eqref{eq:DSbetaDeformation_gen} that generates it.
Integral relations like the chain relation and x-unity were uplifted to their superspace analogs.
We speculate that Osborn's superconformal star integral \eqref{eq:Osborn_original} might play the role of a supersymmetric version of the STR.
However, whether it allows for the construction of commuting transfer matrices is still an open question.
Still, we were able to perform the seminal ``zeroth-order integrability check'' of Zamolodchikov with a positive result, condensed in the determination of the critical coupling \eqref{eq:CritCoup_DS}. 
Furthermore, we were able to perform the analysis in the generality of our newly introduced deformation \eqref{eq:DSbetaDeformation_gen} with the result \eqref{eq:CritCoup_DS_gen}.

There are many directions to expand the findings of this work.
First, the most pressing question is the precise role of Osborn's formula and whether or not it may be used to construct a commuting transfer matrix.
This question is closely related to the construction of a $\mathcal{N}=1$ superconformal R-matrix in the sense of \cite{Chicherin:2012yn}. 
For the one-dimensional case, focusing on the superalgebra $s\ell(2 \vert 1)$, an R-matrix was found in terms of a superspace kernel \cite{Derkachov:2001sx,Derkachov:2005hx,Belitsky:2006cp}.
Second, using Osborn's relation as a tool, one may attempt to repeat many of the computations done over the last years in the fishnet theory to the double-scaled $\beta$-deformation, thereby moving them now somewhat closer to $\mathcal{N}=4$ SYM.
Notable examples are the Basso-Dixon diagrams \cite{Basso:2017jwq,Derkachov:2020zvv,Derkachov:2021rrf}, exact correlators \cite{Grabner:2017pgm,Gromov:2018hut,Basso:2021omx} and the TBA for two-point functions \cite{Basso:2018agi,Basso:2019xay}.
Third, based on superconformal symmetry, one could construct a superfishchain in analogy with \cite{Gromov:2019aku,Iakhibbaev:2023dye}.
Finally, one can construct the Yangian of the superconformal algebra and repeat the Yangian bootstrap program \cite{Chicherin:2017frs,Corcoran:2021gda} for supergraphs in the double-scaled $\beta$-deformation or examine the supergeometries related to the supergraphs \cite{Duhr:2022pch,Duhr:2023eld,Duhr:2024hjf}.

In conclusion, one can repeat the analysis also for other superspaces.
A natural case is the double-scaled $\beta$-deformation of ABJM theory.
This is a three dimensional, $\mathcal{N}=2$ superconformal QFT with a superpotential quartic in the chiral superfields.
In this ``superfishnet'' theory the supergraphs form a regular square lattice \cite{Kade:2024lkc}.

\paragraph{Acknowledgements}
We thank Changrim Ahn for initial collaboration on this project. 
We are thankful to Burkhard Eden for useful discussions. 
Furthermore, we thank the anonymous referees for helpful comments.
This work is funded by the Deutsche Forschungsgemeinschaft (DFG, German Research Foundation) - Projektnummer 417533893/GRK2575 “Rethinking Quantum Field Theory”.

%%%%%%%%%%%%%%%%%%%%%%%%%%%%%%%%%%%%%%%%%%%%%%%%%%%%%%%%%%%%%%%%%%%%%%%%%%%%%%%%%%%%%%%%%%%%%%%%%%%%%%%%%%%%%%%%%%%%%%%%%%%%%%%%%%%%%%%%%%%%%
\appendix

\section{Notations}
\label{sec:Notations}

\subsection{Spinor algebra}
The notation follows \cite{Wess:1992cp} and for the sake of completeness we list important relations here as well.
Small Greek indices of anti-commuting spinors are raised and lowered depending on their chirality as
\begin{equation}
\begin{aligned}[c]
\psi^\alpha = \varepsilon^{\alpha \beta} \psi_{\beta} ,\\
\psi_\alpha = \varepsilon_{\alpha \beta} \psi^{\beta} ,
\end{aligned}
\qquad\qquad
\begin{aligned}[c]
\bar{\psi}^{\dot{\alpha}} = \varepsilon^{\dot{\alpha} \dot{\beta}} \bar{\psi}_{\dot{\beta}} ,\\
\bar{\psi}_{\dot{\alpha}} = \varepsilon_{\dot{\alpha} \dot{\beta}} \bar{\psi}^{\dot{\beta}} ,
\end{aligned}
\end{equation}
with the epsilon tensor squaring to identity as $\varepsilon^{\alpha \beta} \varepsilon_{\beta \gamma} = \delta^\alpha_\gamma$. 
Its non-zero components are $\varepsilon_{21} = \varepsilon^{12} = 1$ and $\varepsilon_{12} = \varepsilon^{21} = -1$.

Spinor bilinears are 
\begin{equation}
\begin{aligned}[c]
\psi \chi &= \left( \psi \chi \right) = \psi^\alpha \chi_\alpha = - \psi_\alpha \chi^\alpha ,\\
\bar{\psi} \bar{\chi} &= \left( \bar{\psi} \bar{\chi} \right) = \bar{\psi}_{\dot{\alpha}} \bar{\chi}^{\dot{\alpha}} = - \bar{\psi}^{\dot{\alpha}} \bar{\chi}_{\dot{\alpha}} = \left( \psi \chi \right)^\dagger ,
\end{aligned}
\qquad\qquad
\begin{aligned}[c]
\psi \sigma^\mu \bar{\chi} = \psi^\alpha \sigma^\mu_{\alpha \dot{\alpha}} \bar{\chi}^{\dot{\alpha}} ,\\
\left( \psi \sigma^\mu \bar{\chi} \right)^\dagger = \chi \sigma^\mu \bar{\psi} ,
\end{aligned}
\end{equation}
and we use the bracket notation $\left( \psi \chi \right) = \left( \chi \psi \right)$ if the index contraction is unclear. 
For coinciding spinors and in particular for spinorial Gra\ss mann numbers we further denote $\theta^2 := \left( \theta \theta \right)$ and $\bar{\theta}^2 := \left( \bar{\theta} \bar{\theta} \right)$ and we have the helpful relations
\begin{equation}
\begin{aligned}[c]
\theta^\alpha \theta^\beta &= - \frac{1}{2} \varepsilon^{\alpha \beta} \theta^2 ,\\
\bar{\theta}^{\dot{\alpha}} \bar{\theta}^{\dot{\beta}} &= \phantom{-} \frac{1}{2} \varepsilon^{\dot{\alpha} \dot{\beta}} \bar{\theta}^2 .
\end{aligned}
\qquad\qquad
\begin{aligned}[c]
\theta \sigma^\mu \bar{\theta}\; \theta \sigma^\nu \bar{\theta} = - \frac{1}{2} \theta^2 \bar{\theta}^2 \eta^{\mu\nu} ,
\end{aligned}
\label{eq:ThetaSquare_ThetaCube}
\end{equation}
Denoting $\bar{\sigma}^{\mu,\dot{\alpha}\alpha} = \varepsilon^{\alpha \beta} \sigma^{\mu}_{\beta\dot{\beta}} \varepsilon^{\dot{\alpha}\dot{\beta}}$, one finds the relations
\begin{equation}
\begin{aligned}[c]
\sigma^{\mu}_{\alpha\dot{\alpha}} \bar{\sigma}_\mu^{\dot{\beta}\beta} = - 2\, \delta_\alpha^\beta \delta_{\dot{\alpha}}^{\dot{\beta}} ,\\
\mathrm{tr} \left[ \sigma^\mu \bar{\sigma}^\nu \right] = - 2\, \eta^{\mu\nu} ,
\end{aligned}
\qquad\qquad
\begin{aligned}[c]
\left[ \sigma^\mu \bar{\sigma}^\nu + \sigma^\nu \bar{\sigma}^\mu \right]_\alpha^{~\beta} = - 2\, \eta^{\mu\nu} \delta_\alpha^\beta ~, \\
\left[ \bar{\sigma}^\mu \sigma^\nu + \bar{\sigma}^\nu \sigma^\mu \right]^{\dot{\alpha}}_{~\dot{\beta}} = - 2\, \eta^{\mu\nu} \delta^{\dot{\alpha}}_{\dot{\beta}} ~ .
\end{aligned}
\end{equation}

\subsection{Berezin integral}
\label{sec:BerezinIntegral}
Integration over the fermionic part of superspace picks out the quadratic component in the Gra\ss mann spinors, such that
\begin{equation}
\int \dd^2 \theta \; \theta^2  =  1
~~ \mathrm{and} ~~
\int \dd^2 \bar{\theta} \; \bar{\theta}^2  =  1~ .
\end{equation}
Furthermore, squares of the Gra\ss mann spinors, appearing in a Berezinian alongside other functions of the integration Gra\ss mann variable, act as delta distributions on fermionic superspace, which we denote by $\delta^{\left(2\right)} \left( \theta_{12} \right) = \theta_{12}^2$ and $\delta^{\left(2\right)} \left( \bar{\theta}_{12} \right) = \bar{\theta}_{12}^2$ with $\theta_{12}= \theta_1 - \theta_2$ and $\bar{\theta}_{12}= \bar{\theta}_1 - \bar{\theta}_2$.
Concerning the mass dimension, Gra\ss mann bilinears have the same values assigned as bosonic coordinates, i.\ e.\ $\left[ \theta^\alpha \right] = \left[ \bar{\theta}^{\dot{\alpha}} \right] = \frac{\left[ x^\mu \right]}{2} = -\frac{1}{2}$.
This implies for the measure of the Berezin integral the mass dimensions $\left[ \dd^2 \theta \right] = \left[ \dd^2 \bar{\theta} \right] = 1$.

The covariant super derivatives and supersymmetry generators are given by
\begin{subequations}
\begin{align}
D_\alpha 
=
\partial_\alpha + \I \sigma^\mu_{\alpha \dot{\alpha}} \bar{\theta}^{\dot{\alpha}} \partial_\mu ~,
\hspace{1cm}
\bar{D}_{\dot{\alpha}} 
=
- \bar{\partial}_{\dot{\alpha}} - \I \theta^\alpha \sigma^\mu_{\alpha \dot{\alpha}} \partial_\mu  ~, \label{eq:SuperCovariantDerivatives}%
\\
Q_\alpha 
=
\partial_\alpha - \I \sigma^\mu_{\alpha \dot{\alpha}} \bar{\theta}^{\dot{\alpha}} \partial_\mu ~,
\hspace{1cm}
\bar{Q}_{\dot{\alpha}} 
=
- \bar{\partial}_{\dot{\alpha}} + \I \theta^\alpha \sigma^\mu_{\alpha \dot{\alpha}} \partial_\mu  ~,
\end{align}
\end{subequations}
with $\partial_\alpha = \frac{\partial}{\partial \theta^\alpha}$, $\bar{\partial}_{\dot{\alpha}} = \frac{\partial}{\partial \bar{\theta}^{\dot{\alpha}}}$ and $\partial_\mu = \frac{\partial}{\partial x^\mu}$.
Note that integration and derivation for Gra\ss mann numbers is equivalent, or in formulas
\begin{subequations}
\begin{align}
\int \dd^2 \theta \; f(\theta)  = \left[ - \frac{1}{4} D^2 f(\theta) \right]_{\theta = 0} ~, &
\hspace{1cm}
\int \dd^2 \bar{\theta} \; f(\bar{\theta})  = \left[ - \frac{1}{4} \bar{D}^2 f(\bar{\theta}) \right]_{\bar{\theta} = 0}  ~,\\
\int \dd^2 \theta\, \dd^2 \bar{\theta} \; f(\theta,\bar{\theta}) 
& = \left[ \frac{1}{16} D^2 \bar{D}^2 f(\theta,\bar{\theta}) \right]_{\theta, \bar{\theta} = 0 } ~.
\end{align}
\end{subequations}

\section{Details of the super-integral calculations}

\subsection{Super chain rule}
\label{subsec:SuperChainRule}
The super chain rules \eqref{eq:ChainRelAntiChiral} and \eqref{eq:ChainRelChiral} can be obtained by direct Gra\ss mann integration of the fermionic coordinates in combination with the bosonic chain rule after performing the Wick rotation,
\begin{equation}
\I \int \dd^Dx_0 
\frac{1}{\left[ x_{10}^2 \right]^{u_1}} 
\frac{1}{\left[ x_{20}^2 \right]^{u_2}}
=
r(D - u_1 - u_2  ,u_1, u_2)
\frac{1}{\left[ x_{12}^2 \right]^{u_1 + u_2 - D/2}} ~.
\label{eq:ChainRel_bosonic}
\end{equation}
Note that the extra factor of $\I$ is due to the fact that throughout this paper we formally stay in Minkowski spacetime, while suppressing the explicit $\I\varepsilon$ prescription for shortness of notation, see also the comment just after \eqref{eq:gen_superPropagator}.

We will present the derivation of \eqref{eq:ChainRelAntiChiral}, the one for \eqref{eq:ChainRelChiral} works analogously.
We start with the integral 
\begin{equation}
\left[
\I \int \dd^4x_0\, \dd^2\bar{\theta}_0
\frac{1}{\left[x_{1\bar{0}}^2 \right]^{u_1}}
\frac{1}{\left[x_{2\bar{0}}^2 \right]^{u_2}}
\right]_{\substack{\theta_0 = 0 \\ \bar{\theta}_{1,2}=0}}
\end{equation}
and represent the fermionic dependency via shift operators, c.\ f.\ \eqref{eq:gen_superPropagator}. 
Afterwards, we can execute the integral over the bosonic subspace by \eqref{eq:ChainRel_bosonic} to find
\begin{equation}
\begin{split}
\int \dd^2\bar{\theta}_0\;
\e^{- 2\I \theta_1 \sigma^\mu \bar{\theta}_0 \partial_{1,\mu}}
\e^{- 2\I \theta_2 \sigma^\nu \bar{\theta}_0 \partial_{2,\nu}}\;
\I \int \dd^4x_0\, 
\frac{1}{\left[x_{10}^2 \right]^{u_1}}
\frac{1}{\left[x_{20}^2 \right]^{u_2}}\\
=
\int \dd^2\bar{\theta}_0\;
\e^{- 2\I \theta_{12} \sigma^\mu \bar{\theta}_0 \partial_{1,\mu}}
\frac{r (4 - u_1 - u_2  ,u_1, u_2)}{\left[ x_{12}^2 \right]^{u_1 + u_2 - 2}} ~.
\end{split}
\end{equation}
We used the fact that $\partial_{2,\mu}$ can be replaced by $-\partial_{1,\mu}$, when acting on a function of $x_{12}$.
The Gra\ss mann integral can now be executed, which amounts to picking the second order in the expansion of the exponential. 
Consecutively using \eqref{eq:ThetaSquare_ThetaCube}, leaves us with 
\begin{equation}
\theta_{12}^2 
\square_1 
\frac{r (4 - u_1 - u_2  ,u_1, u_2)}{\left[ x_{12}^2 \right]^{u_1 + u_2 - 2}}
=
- 4 \,
r (3 - u_1 - u_2  ,u_1, u_2 )
\frac{\theta_{12}^2 }{\left[ x_{12}^2 \right]^{u_1 + u_2 - 1}} ~,
\end{equation}
where we used the functional relation of the gamma function after performing the derivative $\square_{1} \frac{1}{\left[ x_{12}^2 \right]^u} = - 4u (\frac{D}{2} - u -1) \frac{1}{\left[ x_{12}^2 \right]^{u+1}}$.

\subsection{Super x-unity}
\label{subsec:SuperXunity}
The super x-unity relation can be derived from \eqref{eq:ResolutionOfUnity_chiral}, \eqref{eq:3ptFctnWithOneVanishingParameter} and \eqref{eq:3ptFctnOnePointIntegrated} by the following steps:
\begin{subequations}
\begin{align}
\adjincludegraphics[valign=c,scale=0.6]{pictures/xmove/left/LHS/xmove1.pdf} 
&\stackrel{\eqref{eq:ResolutionOfUnity_chiral}}{=}
\lim_{\delta\rightarrow 0} \lim_{\varepsilon\rightarrow 0}
\frac{-1}{4\pi^4 a(\varepsilon)a(3 - \varepsilon)}
\adjincludegraphics[valign=c,scale=0.6]{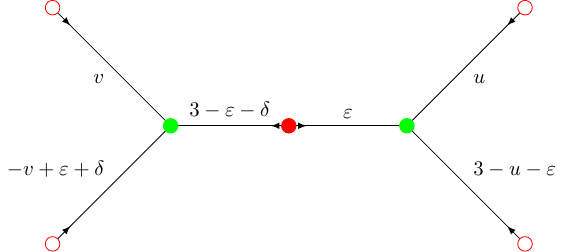}\\
&\stackrel{\eqref{eq:3ptFctnWithOneVanishingParameter}}{=}
\left[
\lim_{\varepsilon\rightarrow 0}
\frac{a(u)a(3-u)}{a(\varepsilon)a(3-\varepsilon)}
\right]
\lim_{\delta\rightarrow 0}
\adjincludegraphics[valign=c,scale=0.6]{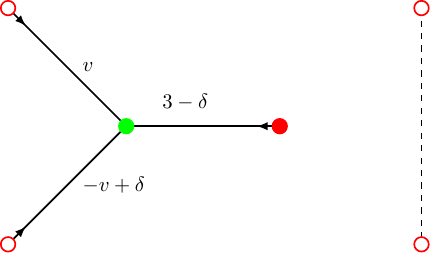}\\
&\stackrel{\eqref{eq:3ptFctnOnePointIntegrated}}{=}
\left[
\lim_{\substack{\varepsilon\rightarrow 0\\\delta\rightarrow 0}}
- 4\pi^4 a(u)a(3-u)
\frac{a(\delta)a(3-\delta)}{a(\varepsilon)a(3-\varepsilon)}
\right]
\adjincludegraphics[valign=c,scale=0.6]{pictures/xmove/left/RHS/xmove7_susy.pdf}\\
&=
- 4\pi^4
a(u)a(3-u) \hspace{0.5cm}
\adjincludegraphics[valign=c,scale=0.6]{pictures/xmove/left/RHS/xmove7_susy.pdf} ~.
\end{align}
\label{eq:SuperXUnity_derivation}
\end{subequations}

\bibliography{Susy_inversion_relations}
\bibliographystyle{OurBibTeX}
\end{document}